\documentclass[conference,compsoc]{IEEEtran}
\usepackage{float}

\ifCLASSOPTIONcompsoc
  \usepackage[nocompress]{cite}
\else
  \usepackage{cite}
\fi
\ifCLASSINFOpdf
  \usepackage[pdftex]{graphicx}
  \graphicspath{{./figures/pngs}}
   \DeclareGraphicsExtensions{.pdf,.jpeg,.png}
\else
\fi
\usepackage{amsmath}
\makeatletter
\renewcommand\IEEEkeywordsname{Keywords}
\def\ps@IEEEtitlepagestyle{
	\def\@oddfoot{\mycopyrightnotice}
	\def\@evenfoot{}
}
\def\mycopyrightnotice{ 
	{\footnotesize } % <--- Change here
	\gdef\mycopyrightnotice{}
}
\begin{document}
\title{A Brief Survey of Current Software Engineering Practices in Continuous Integration and Automated Accessibility Testing}

\author{\IEEEauthorblockN{Parth Sane}
\IEEEauthorblockA{Software Engineering\\
Stratizant Corporation\\
699 Walnut St. Suite 400 \#716\\
Des Moines, IA 50309\\
Email: parth@stratizant.com}
}


% compsoc conferences compared to traditional conferences), use this
%Telephone: (800) 555--1212, Fax: (888) 555--1212}


\maketitle

\begin{abstract}
It's long been accepted that continuous integration (CI) in software engineering increases the code quality of enterprise projects when adhered to by it's practitioners. But is any of that effort to increase code quality and velocity directed towards improving software accessibility accommodations?  What are the potential benefits quoted in literature? Does it fit with the modern agile way that teams operate in most enterprises? This paper attempts to map the current scene of the software engineering effort spent on improving accessibility via continuous integration and it's hurdles to adoption as quoted by researchers. We also try to explore steps that agile teams may take to train members on how to implement accessibility testing and introduce key diagrams to visualize processes to implement CI based accessibility testing procedures in the software development lifecycle.
\end{abstract}

\begin{IEEEkeywords}
	Software Engineering, Accessibility Testing, Continuous Integration, Continuous Delivery.
\end{IEEEkeywords}

% \fi
\IEEEpeerreviewmaketitle

\section{Introduction}
Accessibility testing is very important part of user acceptance testing. There are many scenarios where accessibility testing is performed as described in \cite{eler2018automated} for mobile accessibility testing. \\
In \cite{modelbasedjstesting} a testing framework is proposed where web interfaces using Javascript are tested over a period of time as a long running test. Potentially this could be extended to run long executing accessibility testing. There has been effort by researchers to develop tools to assist public entities like governments to develop more accessible tools\cite{WATforSGgovt}. The interesting part is that they use Continuous Integration (CI) and Continuous Delivery (CD) platforms to help scale up adoption and accessibility testing\cite{WATforSGgovt}.\\
Adopting continuous integration and delivery processes has an overall benefit. This is not solely limited to cost reductions or carrying less technical debt.  Projects adopting continuous integration and automated testing have reported increased number of test cases. There is also reference to enhanced code quality due to more testing or test cases as part of the CI/CD process\cite{8115619}. Zhao et al. in \cite{8115619} find that adopting CI does not increase the complexity of builds. They follow Fowler's "good practices" proposal\cite{fowler_2006}. This means they likely end up spending more time on automated testing after adopting CI practices.\\
Some of the Fowler's principles\cite{fowler_2006} are listed as:
\begin{enumerate}
	\item Maintain a single repository
	\item Automate the build
	\item Make your build self-testing
	\item Everyone commits to the mainline everyday
	\item Every commit should build the mainline on an Integration machine
	\item Fix Broken builds immediately
	\item Keep the build fast
	\item Test in a clone of the Production environment
	\item Make it easy for anyone to get the latest executable
	\item Everyone can see what's happening
	\item Automate Deployment
\end{enumerate}
We can understand the benefits of adopting the same principles for CI on his now influential blog post\cite{fowler_2006} with a detailed comparison.\\
Another interesting part in CI/CD processes is the inclusion of automated testing, as discussed earlier. Are there case studies which explore this in industries or enterprises? Can we further discuss the possibility of accessibility testing with agile\cite{aoyama1998agile} teams?\\
Government mandates and international obligations are strong incentives to drive up adoption too. Vellerman, Nahuis and van der Geest in \cite{velleman2017factors} report that international bodies like the United Nations have legal instruments and agreements like the Convention on the Rights of persons with disabilities\cite{guide2010convention}. These are some of the social factors which drive up adoption as per their report\cite{velleman2017factors}.
\section{Research Questions}
\begin{enumerate}
	\item \textbf{RQ1:} What are some of the challenges to adoption of accessibility testing quoted in literature?\\
	\item \textbf{RQ2:} What are some of the popular software accessibility tools that can be integrated into CI/CD as part of automated tests?
	\item \textbf{RQ3:} What are some of the positive benefits that arise from the implementation of accessibility testing as part of a CI/CD pipeline?
	\item \textbf{RQ4:} Can accessibility testing be incorporated into the agile methodology as part of CI/CD?
	\item \textbf{RQ5:} How can agile teams adopt accessibility testing if they are not familiar with it? 
\end{enumerate}
\section{Adoption of accessibility standards}
Below we are able to examine the reasons cited by researchers why there is acceptance and rejection, due to different reasons with respect to adoption of accessibility testing.
\subsection{Introduction to Accessibility Standards}
Some of the more popular accessibility standards are set by the W3C consortium's WAI initiative\cite{(wai)}. Web accessibility relies on several components that work together as given on \cite{(waistandards)}. Primarily they are:
\begin{enumerate}
	\item \textbf{Web content:} Refers to any part of a website, including text, images, forms, and multimedia, as well as any markup code, scripts, applications, and such.
	\item \textbf{User agents:} Software that people use to access web content, including desktop graphical browsers, voice browsers, mobile phone browsers, multimedia players, plug-ins, and some assistive technologies.
	\item \textbf{Authoring tools:}  Software or services that people use to produce web content, including code editors, document conversion tools, content management systems, blogs, database scripts, and other tools.
\end{enumerate}
Additionally, Yan and Ramachandran \cite{accessibilitymobileapps} provide a quantitative method to calculate the degree of accessibility friendliness. They introduce Inaccessible Element Rate (IAER) and Accessibility Issue Rate (AIR) as two metrics to compute  in order to measure how accessible a website is. Lower IAER and AIR scores indicate better accessibility of an application. These quantitative measures can serve as a benchmark for measuring accessibility. Yan and Ramachandran report in their paper \cite{accessibilitymobileapps} that 67\% of accessibility issues are detected by the IBM Mobile Accessibility Checker. In addition, IAER and AIR scores aim to provide a way to provide quantitative standards that can be adopted by the wider research community and industry. Mars Ballantyne et. al \cite{hawker} also provide a deep study on some accessibility standards and prepare guidelines for testable accessibility standards with a heuristics evaluation process. This heuristics discovery process provides an additional way to judge how accessible a mobile application.
\subsection{Headwinds and Tailwinds to Accessibility Testing adoption}
\subsubsection{Reasons for adopting accessibility testing} ~\\
A lot of value addition can be realized by virtue of making sure software is accessible. From a point of common sense logic, accessible software is not only available to more people but it is a good business case for a commercial venture hawking software of any kind.\\
A study by Jurca et. al\cite{jurca2014integrating} finds surprising cost benefits to adopting accessibility testing for agile scrum teams\cite{jurca2014integrating}. Accessibility testing is cheaper to implement at earlier stages rather than carrying technical debt. Furthermore, making it a part of the process during productization of software makes it better to manage cost on an ongoing basis. It is introduced into this process via automated test cases for accessibility via continuous integration, which can feed into continuous delivery. \\
This makes it a compelling case for a business to integrate accessibility testing into their software development life cycle via having continuous integration as a process that they use for automated testing.

\subsubsection{Reasons for possibly not adopting accessibility testing}~\\
In \cite{10.1145/2207016.2207022} Watanable et. al present that software developers are not always aware of the accessibility guidelines and therefore require tools and evaluation methodologies that help them integrate accessibility as a design and implementation concept into their code base. Another observation is that they are not able to design web interfaces with features like ARIA (Accessible Rich Internet Applications) \cite{w3c-aria} due to implementation level assumptions like dynamic Document Object Model (DOM) modifications of the front end interface that may not be able to work with accessibility tools. For example software like screen-readers that are not very robust may not read dynamically updated DOM in a web page.\\ 
Part of the challenge is to design for accessibility by domain. Notice how the W3C standards are more focused towards web accessibility. The challenge is to have tools that can be targeted to the medium and educating engineers to use. Here 'medium' refers to web, mobile or any other way that a user interacts with software.

\section{Findings of Research Questions}
\subsection{RQ1: What are some of the challenges to adoption of accessibility testing quoted in literature?}
Vellerman, Nahuis and van der Geest in \cite{velleman2017factors} describe that adoption of accessibility technology and standards is often driven by innovation in enterprises. They further go to describe the increased acceptance of technological solutions for accessibility is driven by diffusion of innovation as described by Rogers\cite{rogers2010diffusion}. The pace at which adoption is driven by described by Rogers \cite{rogers2010diffusion} with the following factors:
\begin{enumerate}
	\item \textbf{Relative Advantage:} The perceived competitive edge that is apparent drives how quickly it is adopted.
	\item \textbf{Compatibility:} Is the proposed solution easily adoptable with existing technology? If yes, this is a positive factor for its progress.
	\item \textbf{Complexity:} The extent to which the accessibility standards are understood to be easy or complex to adopt.
	\item \textbf{Observability:} The degree to which adoption and guideline implementation is available to stakeholders.
\end{enumerate}
In companies that are small business, funding for innovation may be limited by the fact that major budget allocations are often for the survival and daily activities like payroll and overhead. Linton et al. \cite{linton2017technology} report as quoted: 
\begin{quote}
``Common but less frequently considered, are small business that change their product offering or underlying process through the adaption of new technology. In some cases, this change is part of a drive to grow sales and/or profitability, whereas in other cases the successful adoption and integration of technology is needed for the survival and sustainability of the firm."
\end{quote}
Thus if accessibility test cases are not an overall priority for the organization then there is little incentive to back it up with real dollars to fund personnel hours required to properly implement it with CI/CD processes. This assumes the fact that they even have a CI/CD pipeline in place and they perform automated test cases.\\
As discussed in the previous section, awareness, tooling and education are some of the major challenges. Watanable's\cite{10.1145/2207016.2207022} comment accurately describes the lack of awareness that software developers often have. As quoted in \cite{10.1145/2207016.2207022}:
\begin{quote}
	 ``Current accessibility evaluation and repair tools present limited crawling capabilities (analysing solely static HTML content) and inability to analyse dynamically generated DOM(Document Object Model) elements, necessary for evaluating RIA accessibility\cite{RIA}."
\end{quote}
 Although there are some hurdles to creating automatic test cases, they are possible to overcome. We are able to make test assertion on dynamically generated and updated contents of webpages as reported in \cite{10.1145/2207016.2207022}.
 
\subsection{RQ2: What are some of the popular software accessibility tools that can be integrated into CI/CD as part of automated tests?}
There are commercial and other free tools available to actually test web interfaces for accessibility issues. Many of these rely on the AXE\cite{deque} rule engine. An example of this tools use case is described in the study by Khwaja \cite{khawaja2020software} which is used to detect accessibility issues in websites of public libraries.\\
Another tool Browserstack\cite{browserstack} provides a guide on how to integrate AXE with their commercial testing product as a Java Selenium\cite{dos2020selenium} test. On the Free and Open Source software side of things, \cite{pa11y} is a popular Python programming language based tool that is integrated or can be incorporated into CI/CD processes. Gitlab, the popular online source code sharing and collaboration platform uses Pa11y\cite{pa11y} as a part of a simple CI job template\cite{gitlab-accessibility-template}. That CI job template spits out various accessibility violations, warnings and notices to an output file. \\
Using Gitlab's Pipelines feature for repositories, it becomes possible to test code without any special infrastructure or setup from the part of developers and User Interface \& Experience designers(UI/UX).
\subsection{RQ3: What are some of the positive benefits that arise from the implementation of accessibility testing as part of a CI/CD pipeline?}
Projects that use CI/CD pipelines as part of their process have test automation as a prevalent practice as compared to projects that don't\cite{cruz2019attention}. As we already suspect, a non-zero number of tests in a project are correlated with higher code quality.\\
As per Hilton et. al \cite{10.1145/2970276.2970358} CI/CD has proven to be beneficial and to have improved benefits when employed with projects having automated test cases.They emphasize that CI helps in catching bugs not locating them. We often get an important error message buried in CI logs which may be difficult to pour over manually. These may be messages from tools like AXE \cite{deque} as seen in the example\cite{gitlab-accessibility-template} that Gitlab provides for deploying a CI/CD solution.\\
Quantitatively, we can get more data and learn more from metrics gathered from tooling as a part of deployed CI/CD solutions. Gitlab's solution is a good start to customize what kind of metrics can be collected to get insights.
\subsection{RQ4: Can accessibility testing be incorporated into the agile methodology as part of CI/CD?}
\begin{figure}
\centering
\includegraphics[scale=0.2]{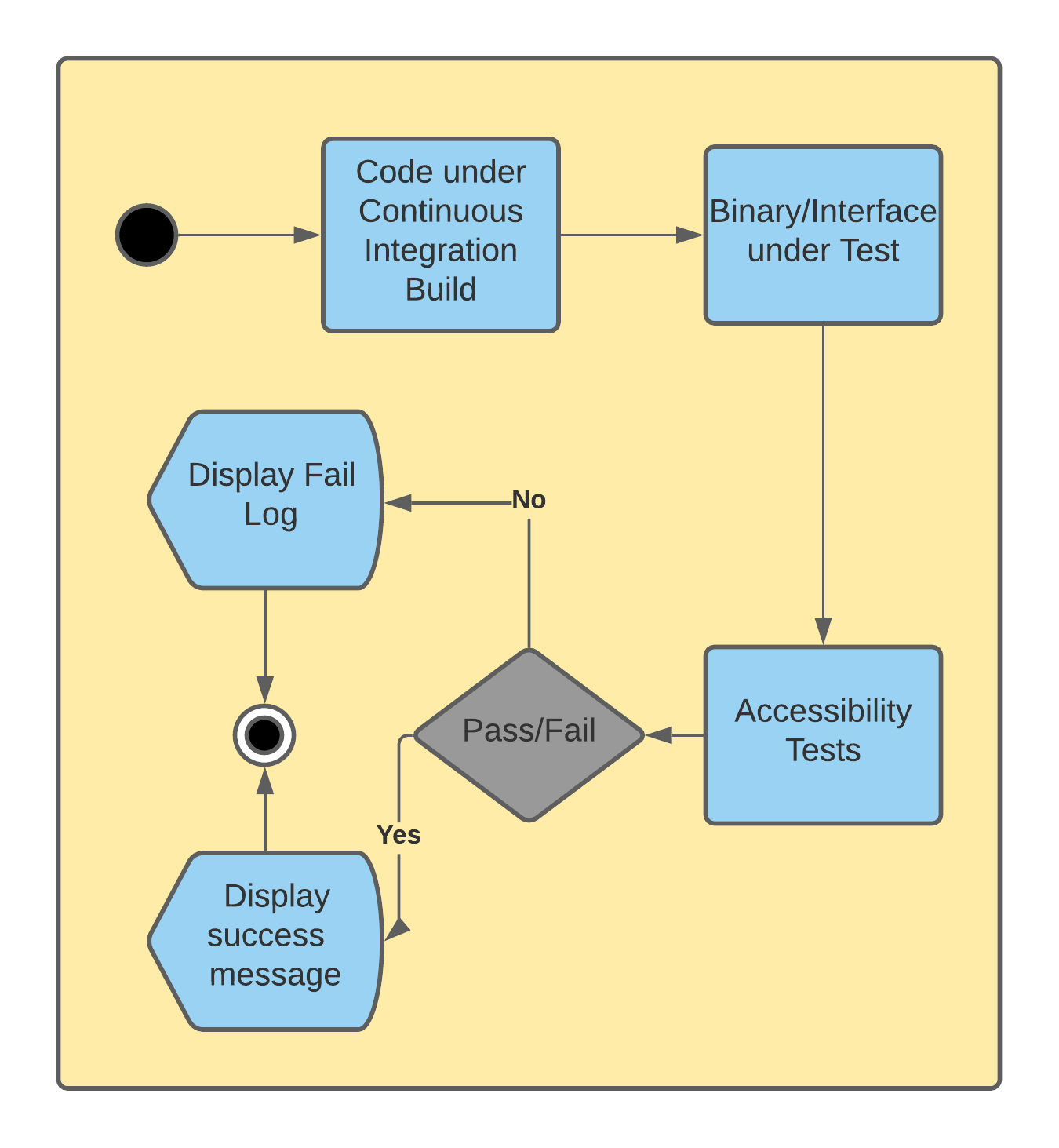}
\caption{Automated Accessibility Tests as a part of Continuous Integration}
\label{AutoAccTestFig}
\end{figure}

Accessibility tests are or can be a part of standard test automation packages as seen in \cite{gitlab-accessibility-template}. As seen in Fig.\ref{AutoAccTestFig}, automated accessibility tests can be included as part of the software development life-cycle. The Agile methodology often emphasizes a build fast and try to break often kind of testing philosophy. \\
The software development team can incorporate their accessibility tests with their regular unit tests for their code. This will help unify concepts of automated accessibility tests that were not the approach earlier, but now can be as freshly proposed in Fig.\ref{AutoAccTestFig}.

Fig. \ref{AutoAccTestFig} can be explained briefly with the following steps:
\begin{enumerate}
	\item Code Enters Build in CI
	\item Binary/Code Interface in Test
	\item Automated accessibility tests using AXE\cite{deque} integration with Gitlab or other CI Build Systems
	\item Depending on results generate pass or fail results
	\item Output log to iteratively fix outstanding issues.
\end{enumerate}
\subsection{RQ5: How can agile teams adopt accessibility testing if they are not familiar with it? }
\begin{figure}[H]
	\centering
	\includegraphics[scale=0.64]{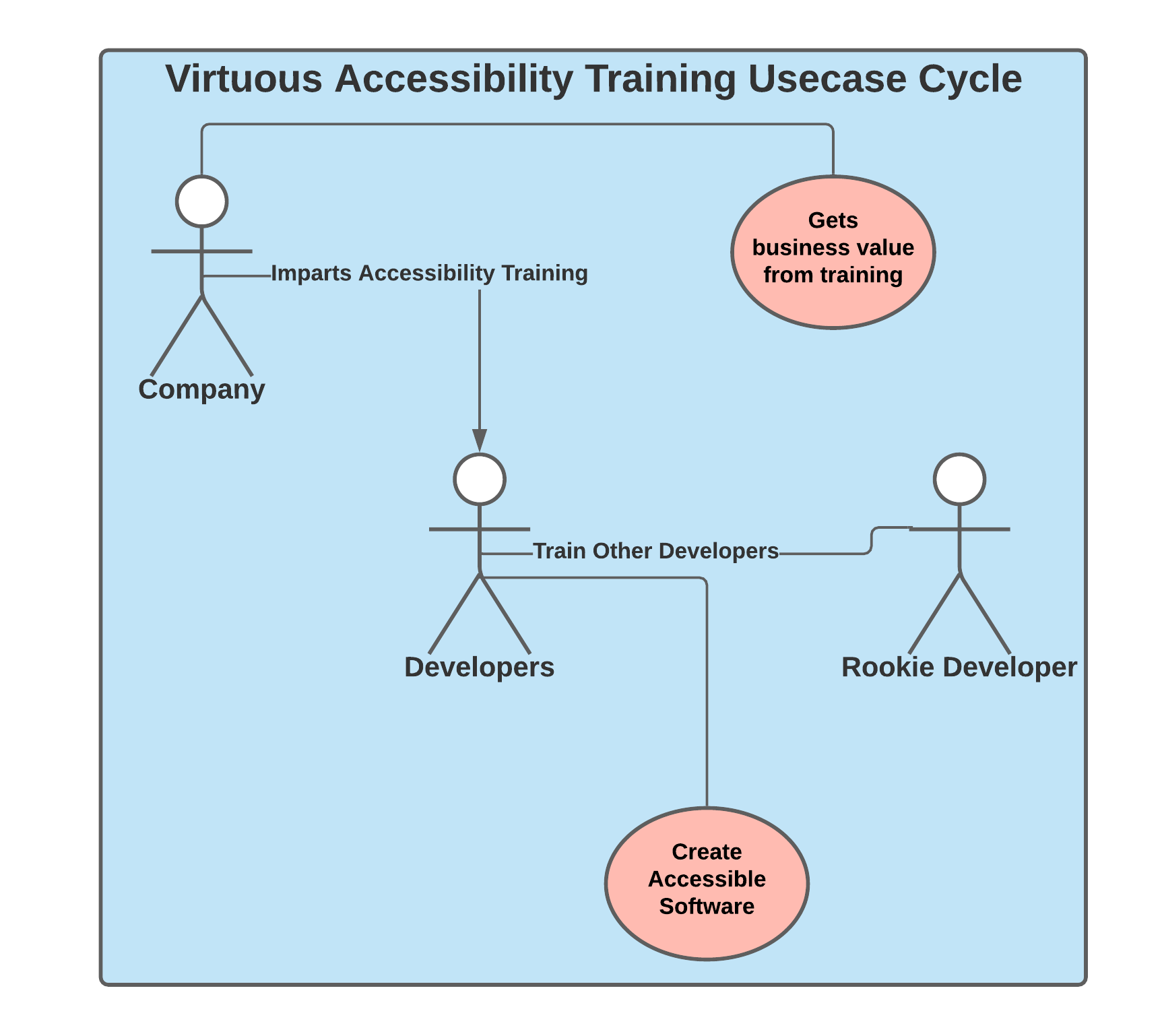}
	\caption{Virtuous Accessibility Training Cycle Pseudo-use-case diagram}
	\label{VirtuousAcc}
\end{figure}

Sverdup and Sand\cite{sverdrup2018accessibility} describe some of the challenges faced by teams adopting accessibility. They concur with other studies\cite{10.1145/2207016.2207022} that cite accessibility in software products is an underdeveloped area.
Sverdup reports that software developers generally tend to do worse in implementing accessibility. The survey conducted reports that there was a general feeling that implementing accessibility testing was outside their job responsibilities.\\
Some tools suggested by Sverdup to make it easier to meet rapid sprint deadlines is to use physical tools so that developers can empathize with users affected by physical constraints. Some of them are:
\begin{enumerate}
	\item Cambridge Simulation Glasses: Devices that blurs the testers vision, allowing the tester to experience the tested software product from the perspective of the user with reduced vision.
	\item Dyslexia Simulator/SiteImprove test tool
	\item Personas: Developers test the software from a point of view of imagined users with disabilities.
	\item Screen-readers: These are tools that users with reduced or no vision utilize to browse the internet. 
\end{enumerate}
Serious training effort is required by enthusiastic middle management to train developers and test engineers to upgrade their skills and incorporate accessibility testing into the software development and delivery process. Some of the model curricula in literature are the Adobe model and the Teach Access model. Some of the features are as given in \cite{ladner2017teaching}:
\begin{enumerate}
	\item Adobe model:
	\begin{itemize}
		\item 8 hours of engineering training
		\item Roles/States/Properties/Events
		\item Keyboard accessibility
		\item Testing with assistive technology
		\item Desktop APIs (Windows/macOS)
		\item Mobile APIs (iOS/Android)
	\end{itemize}
	\item Teach Access Model
	\begin{itemize}
		\item Developing a train-the-trainer session
		\item 2 hour lecture with supporting material
		\item Disability awareness/societal context
		\item Roles/States/Properties/Events
		\item Common barriers to using IT
		\item Common assistive technologies
		\item Best practices for design/engineering
		\item Applied techniques
	\end{itemize}
\end{enumerate} 
We can read more about the above given models in Ladner and May\cite{ladner2017teaching}. These are some of the good ways proposed in literature to teach accessibility. These can be extended to a more corporate environment and shortened to accommodate modern rapid timelines. Additionally the following pseduo-usecase diagram is proposed as Fig.\ref{VirtuousAcc}. It explains and elaborates more on how accessibility education can happen in the corporate environment.
\underline{$  $}

\section{Limitations}
The main goal of this study and survey is to research the state of accessibility testing in continuous integration.  As such the most evident limitation of this study is the lack of quantitative data presented. Qualitative, rather than quantitative information is presented in an effort to demonstrate the various practical and operational aspects of implementing continuous integration driven automated accessibility testing. Further ways in which this research could be improved or extended is by doing a case study for an organization. A pre and post-adoption interview of developers, sprint metrics and customer feedback surveys can be a great case study to measure the practical issues and benefits in implementing accessibility testing.\\
Additionally, suggestions presented herein are not exhaustive by all means, since new standards, technologies and tools are being agreed upon by the community from time to time. But this is a good review of the overall state of the practice with respect to automated accessibility tests in a CI/CD process.
\section{Conclusion and Key Contributions}
The main takeaway from this study is the big benefits to businesses and users in need to accessible software. The business gets to address an under-served market. Accessible software benefits users and serves a symbiotic need in the industry. Furthermore, we have seen in numerous publications cited in this paper that automated accessibility testing is a feature that is a need of the hour. We also benefit from the agile sprint cycles that developers publish their code in by taking advantage of Continuous Integration and Continuous Delivery process methodologies. Increased code quality is associated with increased adoption of automated tests in a CI/CD environment. Fig. \ref{AutoAccTestFig} and \ref{VirtuousAcc} introduced in this paper provide a starting template for software teams to visualize processes to be adopted as part of their sprint cycles. These two figures are two of the key contributions of this paper.
\\
This blends in well with existing enthusiasm with Software Developers that are ready and eager to learn how to implement and develop accessible software as this can amount to a positive impact on the lives of those who depend on it. This responsibility falls upon enterprises and businesses to educate and train these workers by enabling them in adding value to software that they in turn sell.


\ifCLASSOPTIONcompsoc
  \section*{Acknowledgments}
\else
  \section*{Acknowledgment}
\fi

I would like to thank Brad Hoyt for his support and interest in supporting research at Stratizant Corporation.



%\bibliographystyle{IEEEtran}
\bibliographystyle{IEEEtran}
\bibliography{bibi}

\begin{thebibliography}{10}
\providecommand{\url}[1]{#1}
\csname url@samestyle\endcsname
\providecommand{\newblock}{\relax}
\providecommand{\bibinfo}[2]{#2}
\providecommand{\BIBentrySTDinterwordspacing}{\spaceskip=0pt\relax}
\providecommand{\BIBentryALTinterwordstretchfactor}{4}
\providecommand{\BIBentryALTinterwordspacing}{\spaceskip=\fontdimen2\font plus
\BIBentryALTinterwordstretchfactor\fontdimen3\font minus
  \fontdimen4\font\relax}
\providecommand{\BIBforeignlanguage}[2]{{%
\expandafter\ifx\csname l@#1\endcsname\relax
\typeout{** WARNING: IEEEtran.bst: No hyphenation pattern has been}%
\typeout{** loaded for the language `#1'. Using the pattern for}%
\typeout{** the default language instead.}%
\else
\language=\csname l@#1\endcsname
\fi
#2}}
\providecommand{\BIBdecl}{\relax}
\BIBdecl

\bibitem{eler2018automated}
M.~M. Eler, J.~M. Rojas, Y.~Ge, and G.~Fraser, ``Automated accessibility
  testing of mobile apps,'' in \emph{2018 IEEE 11th International Conference on
  Software Testing, Verification and Validation (ICST)}.\hskip 1em plus 0.5em
  minus 0.4em\relax IEEE, 2018, pp. 116--126.

\bibitem{modelbasedjstesting}
\BIBentryALTinterwordspacing
G.~Brajnik, C.~Pighin, and S.~Fabbro, ``Model-based automated accessibility
  testing,'' in \emph{Proceedings of the 17th International ACM SIGACCESS
  Conference on Computers \& Accessibility}, ser. ASSETS '15.\hskip 1em plus
  0.5em minus 0.4em\relax New York, NY, USA: Association for Computing
  Machinery, 2015, p. 319–320. [Online]. Available:
  \url{https://doi.org/10.1145/2700648.2811357}
\BIBentrySTDinterwordspacing

\bibitem{WATforSGgovt}
\BIBentryALTinterwordspacing
Z.~Y. Lim, J.~M. Chua, K.~Yang, W.~S. Tan, and Y.~Chai, ``Web accessibility
  testing for singapore government e-services,'' in \emph{Proceedings of the
  17th International Web for All Conference}, ser. W4A '20.\hskip 1em plus
  0.5em minus 0.4em\relax New York, NY, USA: Association for Computing
  Machinery, 2020. [Online]. Available:
  \url{https://doi.org/10.1145/3371300.3383353}
\BIBentrySTDinterwordspacing

\bibitem{8115619}
Y.~{Zhao}, A.~{Serebrenik}, Y.~{Zhou}, V.~{Filkov}, and B.~{Vasilescu}, ``The
  impact of continuous integration on other software development practices: A
  large-scale empirical study,'' in \emph{2017 32nd IEEE/ACM International
  Conference on Automated Software Engineering (ASE)}, 2017, pp. 60--71.

\bibitem{fowler_2006}
\BIBentryALTinterwordspacing
M.~Fowler, ``Continuous integration,'' May 2006. [Online]. Available:
  \url{https://www.martinfowler.com/articles/continuousIntegration.html}
\BIBentrySTDinterwordspacing

\bibitem{aoyama1998agile}
M.~Aoyama, ``Agile software process and its experience,'' in \emph{Proceedings
  of the 20th international conference on Software engineering}.\hskip 1em plus
  0.5em minus 0.4em\relax IEEE, 1998, pp. 3--12.

\bibitem{velleman2017factors}
E.~M. Velleman, I.~Nahuis, and T.~van~der Geest, ``Factors explaining adoption
  and implementation processes for web accessibility standards within
  egovernment systems and organizations,'' \emph{Universal access in the
  information society}, vol.~16, no.~1, pp. 173--190, 2017.

\bibitem{guide2010convention}
T.~Guide, ``The convention on the rights of persons with disabilities,'' 2010.

\bibitem{(wai)}
\BIBentryALTinterwordspacing
W.~W. A.~I. (WAI). [Online]. Available: \url{https://www.w3.org/WAI/}
\BIBentrySTDinterwordspacing

\bibitem{(waistandards)}
\BIBentryALTinterwordspacing
------, ``Accessibility principles.'' [Online]. Available:
  \url{https://www.w3.org/WAI/fundamentals/accessibility-principles/\#standards}
\BIBentrySTDinterwordspacing

\bibitem{accessibilitymobileapps}
\BIBentryALTinterwordspacing
S.~Yan and P.~G. Ramachandran, ``The current status of accessibility in mobile
  apps,'' \emph{ACM Trans. Access. Comput.}, vol.~12, no.~1, Feb. 2019.
  [Online]. Available: \url{https://doi.org/10.1145/3300176}
\BIBentrySTDinterwordspacing

\bibitem{hawker}
\BIBentryALTinterwordspacing
M.~Ballantyne, A.~Jha, A.~Jacobsen, J.~S. Hawker, and Y.~N. El-Glaly, ``Study
  of accessibility guidelines of mobile applications,'' in \emph{Proceedings of
  the 17th International Conference on Mobile and Ubiquitous Multimedia}, ser.
  MUM 2018.\hskip 1em plus 0.5em minus 0.4em\relax New York, NY, USA:
  Association for Computing Machinery, 2018, p. 305–315. [Online]. Available:
  \url{https://doi.org/10.1145/3282894.3282921}
\BIBentrySTDinterwordspacing

\bibitem{jurca2014integrating}
G.~Jurca, T.~D. Hellmann, and F.~Maurer, ``Integrating agile and user-centered
  design: a systematic mapping and review of evaluation and validation studies
  of agile-ux,'' in \emph{2014 Agile Conference}.\hskip 1em plus 0.5em minus
  0.4em\relax IEEE, 2014, pp. 24--32.

\bibitem{10.1145/2207016.2207022}
\BIBentryALTinterwordspacing
W.~M. Watanabe, R.~P.~M. Fortes, and A.~L. Dias, ``Using acceptance tests to
  validate accessibility requirements in ria,'' in \emph{Proceedings of the
  International Cross-Disciplinary Conference on Web Accessibility}, ser. W4A
  '12.\hskip 1em plus 0.5em minus 0.4em\relax New York, NY, USA: Association
  for Computing Machinery, 2012. [Online]. Available:
  \url{https://doi.org/10.1145/2207016.2207022}
\BIBentrySTDinterwordspacing

\bibitem{w3c-aria}
\BIBentryALTinterwordspacing
W3C. [Online]. Available: \url{https://www.w3.org/TR/wai-aria-1.1/}
\BIBentrySTDinterwordspacing

\bibitem{rogers2010diffusion}
E.~M. Rogers, \emph{Diffusion of innovations}.\hskip 1em plus 0.5em minus
  0.4em\relax Simon and Schuster, 2010.

\bibitem{linton2017technology}
J.~D. Linton and G.~T. Solomon, ``Technology, innovation, entrepreneurship and
  the small business—technology and innovation in small business,''
  \emph{Journal of small business management}, vol.~55, no.~2, pp. 196--199,
  2017.

\bibitem{RIA}
C.~Velasco, D.~Denev, D.~Stegemann, and Y.~Mohamad, ``A web compliance
  engineering framework to support the development of accessible rich internet
  applications,'' 01 2008, pp. 45--49.

\bibitem{deque}
\BIBentryALTinterwordspacing
Deque. [Online]. Available: \url{https://www.deque.com/axe/}
\BIBentrySTDinterwordspacing

\bibitem{khawaja2020software}
P.~Khawaja, ``A software tool-based accessibility assessment of public library
  websites in the united states,'' 2020.

\bibitem{browserstack}
\BIBentryALTinterwordspacing
BrowserStack, ``Run accessibility tests with automate using axe library:
  Browserstack docs.'' [Online]. Available:
  \url{https://www.browserstack.com/docs/automate/selenium/accessibility-testing}
\BIBentrySTDinterwordspacing

\bibitem{dos2020selenium}
J.~P.~R. dos Santos, K.~P. da~Silva, B.~P.~G. Gon{\c{c}}alves, J.~S.
  de~Souza~Pinheiro, J.~M.~L. de~Oliveira, and D.~B. de~Alencar, ``Selenium as
  a free tool to test for java web application,'' \emph{International Journal
  of Advanced Engineering Research and Science}, vol.~7, no.~4, 2020.

\bibitem{pa11y}
\BIBentryALTinterwordspacing
``Pa11y.'' [Online]. Available: \url{https://pa11y.org/}
\BIBentrySTDinterwordspacing

\bibitem{gitlab-accessibility-template}
\BIBentryALTinterwordspacing
``lib/gitlab/ci/templates/verify/accessibility.gitlab-ci.yml · master ·
  gitlab.org / gitlab.'' [Online]. Available:
  \url{https://gitlab.com/gitlab-org/gitlab/blob/master/lib/gitlab/ci/templates/Verify/Accessibility.gitlab-ci.yml}
\BIBentrySTDinterwordspacing

\bibitem{cruz2019attention}
L.~Cruz, R.~Abreu, and D.~Lo, ``To the attention of mobile software developers:
  guess what, test your app!'' \emph{Empirical Software Engineering}, vol.~24,
  no.~4, pp. 2438--2468, 2019.

\bibitem{10.1145/2970276.2970358}
\BIBentryALTinterwordspacing
M.~Hilton, T.~Tunnell, K.~Huang, D.~Marinov, and D.~Dig, ``Usage, costs, and
  benefits of continuous integration in open-source projects,'' in
  \emph{Proceedings of the 31st IEEE/ACM International Conference on Automated
  Software Engineering}, ser. ASE 2016.\hskip 1em plus 0.5em minus 0.4em\relax
  New York, NY, USA: Association for Computing Machinery, 2016, p. 426–437.
  [Online]. Available: \url{https://doi.org/10.1145/2970276.2970358}
\BIBentrySTDinterwordspacing

\bibitem{sverdrup2018accessibility}
N.~J.~S. Sverdrup, ``Accessibility testing in agile software development,''
  Master's thesis, 2018.

\bibitem{ladner2017teaching}
R.~E. Ladner and M.~May, ``Teaching accessibility,'' in \emph{Proceedings of
  the 2017 ACM SIGCSE Technical Symposium on Computer Science Education}, 2017,
  pp. 691--692.

\end{thebibliography}

\end{document}